\documentclass{article}
\usepackage{caption}
\usepackage{subcaption}
\usepackage{spconf,amsmath,graphicx}
\usepackage{amsfonts}
\usepackage{hyperref}
\usepackage{todonotes}
\usepackage{url}

\usepackage{pgfplots}
\usepackage{pgfplotstable}
\pgfplotsset{compat=1.16}

\newcommand\nnfootnote[1]{%
  \begin{NoHyper}
  \renewcommand\thefootnote{}\footnote{#1}%
  \addtocounter{footnote}{-1}%
  \end{NoHyper}
}

\title{Optimizing Short-Time Fourier Transform Parameters via Gradient Descent}
\name{An Zhao$^{\sharp}$ \qquad Krishna Subramani$^{\sharp}$ \qquad Paris Smaragdis$^{\sharp,\flat}$}
\address{$^{\sharp}$University of Illinois at Urbana-Champaign, $^{\flat}$Adobe Research}

\usepackage{hyperref}
\begin{document}
\ninept
\maketitle
\begin{abstract}
The Short-Time Fourier Transform (STFT) has been a staple of signal processing, often being the first step for many audio tasks. A very familiar process when using the STFT is the search for the best STFT parameters, as they often have significant side effects if chosen poorly. These parameters are often defined in terms of an integer number of samples, which makes their optimization non-trivial. In this paper we show an approach that allows us to obtain a gradient for STFT parameters with respect to arbitrary cost functions, and thus enable the ability to employ gradient descent optimization of quantities like the STFT window length, or the STFT hop size. We do so for parameter values that stay constant throughout an input, but also for cases where these parameters have to dynamically change over time to accommodate varying signal characteristics.
\end{abstract}

\nnfootnote{Code: \texttt{https://github.com/SubramaniKrishna/STFTgrad}}

\section{Introduction}
With the advent of deep learning we have seen a dramatic shift in signal processing towards incorporating neural net-like learning (e.g. with Differentiable DSP research \cite{ddsp}).  Although many parts of signal processing fit well into that framework, some parameters that we often use are not as easy to optimize.  This usually includes parameters defined in terms of samples, such as frame sizes, hop sizes, etc. The Short-Time Fourier Transform \cite{Allen} is a prime example of this.  Previous work has focused on automatically finding optimal STFT parameters, e.g. in using dynamic programming to obtain the best window positions \cite{mergewindows}, or to detect signal non-stationarity to adjust the analysis parameters \cite{stationarity}.  However, most of that work is based on heuristics and local search, and is not compatible with gradient descent-style optimization that can be used to jointly optimize entire end-to-end systems.

Here, we propose two approaches for optimizing STFT parameters via gradient descent.  We show that these can be used for any appropriately defined loss function, and can be easily incorporated in larger systems.

\section{An optimizable STFT}
As is well known, due to the time/frequency trade-off, picking the wrong STFT window size can result in increased smearing across the time or frequency axis, which in turn creates a poor representation of the data. Having the wrong parameters can not only result in a non-legible transform, but also provide a poor feature representation for further processing. Here we will present a formulation of the STFT that will allow us to directly optimize parameters such as the window length, and optimize it with respect to an arbitrary differentiable loss function.

For an input signal $x[t]$, we will define the STFT analysis as:
\begin{equation}
F_{W}[m, k] = \sum_{n = -N/2}^{N/2}{x[m+n]W_m[n]}e^{-j \frac{2\pi}{N} k n}
\label{stft}
\end{equation}
where $m$ is the sample position, $W_m$ is the analysis window function centered at the $m$-th sample, and $F_{W}[m, k]$ is the resulting transform at time $m$ and frequency $k$.  For constant-sized windows and hop sizes, the window $W_m$ is the same for all $m$ (e.g. a Hann window) and the STFT will sample $W_m$ at fixed intervals.  We use however this notation to facilitate the use of windows whose shape is dependent on $m$, as we will use in later sections.

Using this definition, will consider two distinct cases, one in which we are trying to estimate STFT parameters that are constant throughout the analysis (i.e. a fixed hop size, window size, or window shape), and later on the case where the STFT parameters are dynamically changing in order to adapt to the input signal.  We start with the former since it is an easier formulation that can help lead to the next one.

\section{Optimizing for constant STFT parameters}
Here we describe how we can optimize the STFT assuming the window parameters are constant throughout the transform (e.g. using a constant size transform throughout the duration of the signal). We will outline the steps for obtaining a gradient for integer parameters like the window size, and will demonstrate this using a sparsity cost function.

Traditionally STFT uses integer window sizes and integer hop sizes. The window function, as defined in equation \ref{stft}, is a fixed function of the window size. If we wish to optimize using gradient descent, the STFT is a problem since the involved variables are not continuous. By using an underlying continuous variable to derive both the window function and the window size, we can make the STFT parameters differentiable, even though the computed sizes remain discrete.  In the case of the STFT this is relatively straightforward as shown below.

In order to obtain a meaningful differentiable setup, we can define $W_m[n]$ in equation \ref{stft} to use a Gaussian window function:
\begin{equation}\label{gaussian_window}
W_m[n] = \exp\left[-\left( \frac{n}{2\sigma}\right) ^2\right]
\end{equation}
Note that by doing this we do not make a direct use of the window length, we instead use the continuous parameter $\sigma$ as a proxy. Since the value of this window is effectively zero for large values of $|n|$, when computing the transform that uses this window we can safely truncate the window to zero for, e.g., $|n| > 3\sigma$ and assume that it has a length of $N = \left \lfloor{6 \sigma}\right \rfloor$ samples (which is the non-zero region between $-3\sigma$ and $3\sigma$). By computing the results using the truncated window, but optimizing with respect to the infinite one we can optimize the effective DFT length of the transform. In the examples here we fix the hop size to $50\%$ of window length, but that can be changed to any desired value.

\subsection{Optimizing for sparsity}
As an illustrative example we will consider the case where we wish to find STFT parameters that result in the most sparse STFT magnitudes.  This is often a desirable property in time frequency analysis \cite{Gardner,costa2019sparse,wolfe} and will allow us to show how we can take a loss function and directly optimize an STFT parameter using gradient descent.

We will evaluate the sparsity of the transform using the following measure:
\begin{equation}
C[m,W] = \frac{c_4(m, W)}{c_2(m, W)^2}
\end{equation}
which is a measure of \textit{concentration} as defined in \cite{concentration} and \cite{LocalKurtosis} that is derived from the kurtosis (itself a measure of sparsity). In the equation above, $c_2(m,W)$ and $c_4(m,W)$ are $\ell_2$ and $\ell_4$ norms of each time slice of the STFT:
\begin{equation}
\begin{split}
    c_p(m,W) = \sum_{k = 0}^{N}\left| F_{W}[m, k] \right| ^p 
%    c_4(m,W) = \sum_{k = 0}^{N} \mid F_{W}[m, k] \mid ^4\
\end{split}
\end{equation}
In practice, we will sample $C[m,W]$ under the constraint that the windows have reasonable overlap, and will then sum their concentration as the final measure.  Therefore the sparsity loss function over the entire input with the proposed window function is defined as:
\begin{equation}
\mathcal{L}_S = \frac{\sum_{i = 0}^{\frac{M}{3\sigma}}c_4(3\sigma i,W)}{\sum_{i = 0}^{\frac{M}{3\sigma}}c_2(3\sigma i,W)^2}
\end{equation}
And since the entire analysis is now differentiable, we can easily propagate gradients and optimize $\sigma$ to maximize $\mathcal{L}_S$. By doing so we can find the sparsest representation by effectively adjusting the transform's window length.
\begin{figure}[ht]
\centering
\includegraphics[width=0.48\textwidth]{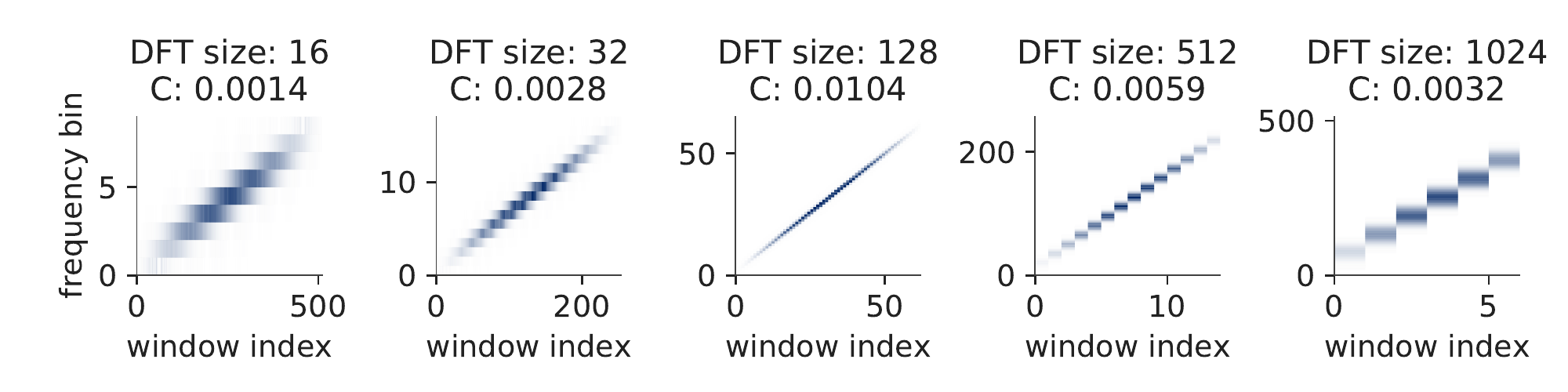}
\caption{Spectrograms of different STFT sizes with their corresponding concentration. Our approach will automatically obtain the sparsest representation (middle plot) via gradient descent, instead of a direct search.}
\label{f1}
\end{figure}
Applying this on a various signals reliably results in a window length that creates the most appropriate STFT representation, for example the middle plot in Figure \ref{f1}. Of course, in this particular case, a direct search of the optimal window size would be faster and easier to perform. In order to demonstrate a more realistic use of this idea we will present it within a more complex context in the next section.

\subsection{Classification Experiments}
We will now examine a more involved case in which we want to tune the STFT parameters in order to optimize a subsequent estimation that will be jointly optimized.  We will do so in a simple sound classification setting.  Our goal this time is to find the optimal STFT parameters in order to optimize a classifier's ability to discern between two sound classes.  Instead of the sparsity measure, we will now use the classifier's loss as the cost function and we will be optimizing the size of the STFT windows as well as the classifier parameters simultaneously.

More formally, we will use the same STFT formulation with the Gaussian window as in the previous section, and a classifier function $K(\cdot)$ that operates on each frame $F_W[m]$ of the STFT and provides a set of class predictions $z_i[m] = K( F_W[m])$ for each class $i$ and time $m$.  For our experiments ahead, we will consider a simple linear classifier mapping from the input dimension (appropriately zero padded to ensure a constant input dimension to the classifier), followed by a softmax to give output $z \, \in \, \mathbb{R}^{2}_{+}$.  We can write a loss function that describes the accuracy of the classifier, and we will also add to it a regularizing term to avoid very small DFT sizes to ensure efficient processing.  The overall loss then becomes:
%More formally, let $T_1(\cdot)$ represent our ``differentiable'' Gaussian Window STFT (as defined in \autoref{gaussian_window}, parameterized by its variance $\sigma^2$), and $K(\cdot)$ represent a classifier. If our classifier loss is $\mathcal{L}_C = f[K(T_1(x)),l(x)]$ for input $x$ and its label $l(x)$, we want to find the optimal STFT window size $N$, along with the optimal parameters for $K(\cdot)$, which minimizes $\mathcal{L}_C$. For our experiments ahead, we will consider a simple linear classifier mapping from the input dimension (appropriately zero padded to ensure a constant input dimension to the classifier), followed by a softmax to give output $z \, \in \, \mathbb{R}^{2}_{+}$.  An observation to make is that if the window size is too small, there will be a lot of processing involved (because of a large number of frames). Furthermore, very small windows might not capture the spectral content of the signal well enough to discriminate them. Thus, we modify the loss by adding a regularizing term which penalizes small windows. Our final network loss is then defined as:
\begin{equation}
\mathcal{L}_C = -\sum_{\forall m}\sum_i t_i[m] \log(z_i[m]) + \frac{\lambda}{\sigma}
\end{equation}
which is the typical cross-entropy loss ($t_i[m]$ being the ground truth and $z_i[m]$ being the network output) with an extra term that penalizes small window sizes.  The constant $\lambda$ defines the strength of the regularizer (which in this experiment is set to 0.1). Using this loss as a guide, we want to find the optimal STFT window size $N$, along with the optimal parameters for the classifier simultaneously.

\begin{figure}[ht]
\centering
\includegraphics[width=0.48\textwidth]{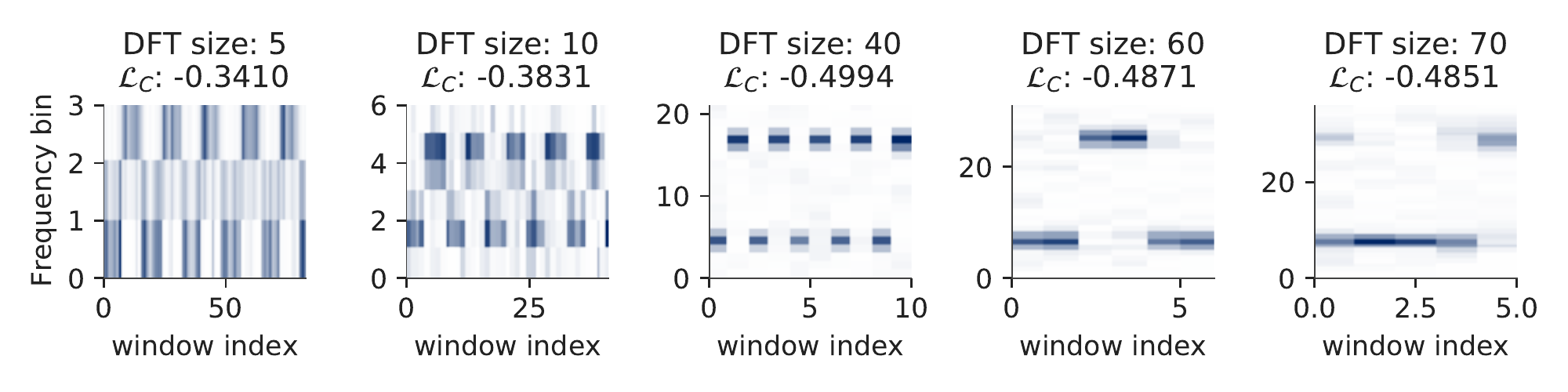}
\caption{Spectrograms of input $x[m]$ for different STFT window size $N$ along with their corresponding $\mathcal{L}_C$. Note how for a DFT size of 40 we get the cleanest representation.}
\label{fig::class_spec}
\end{figure}

\begin{figure}
	\centering
\includegraphics[width=0.48\textwidth]{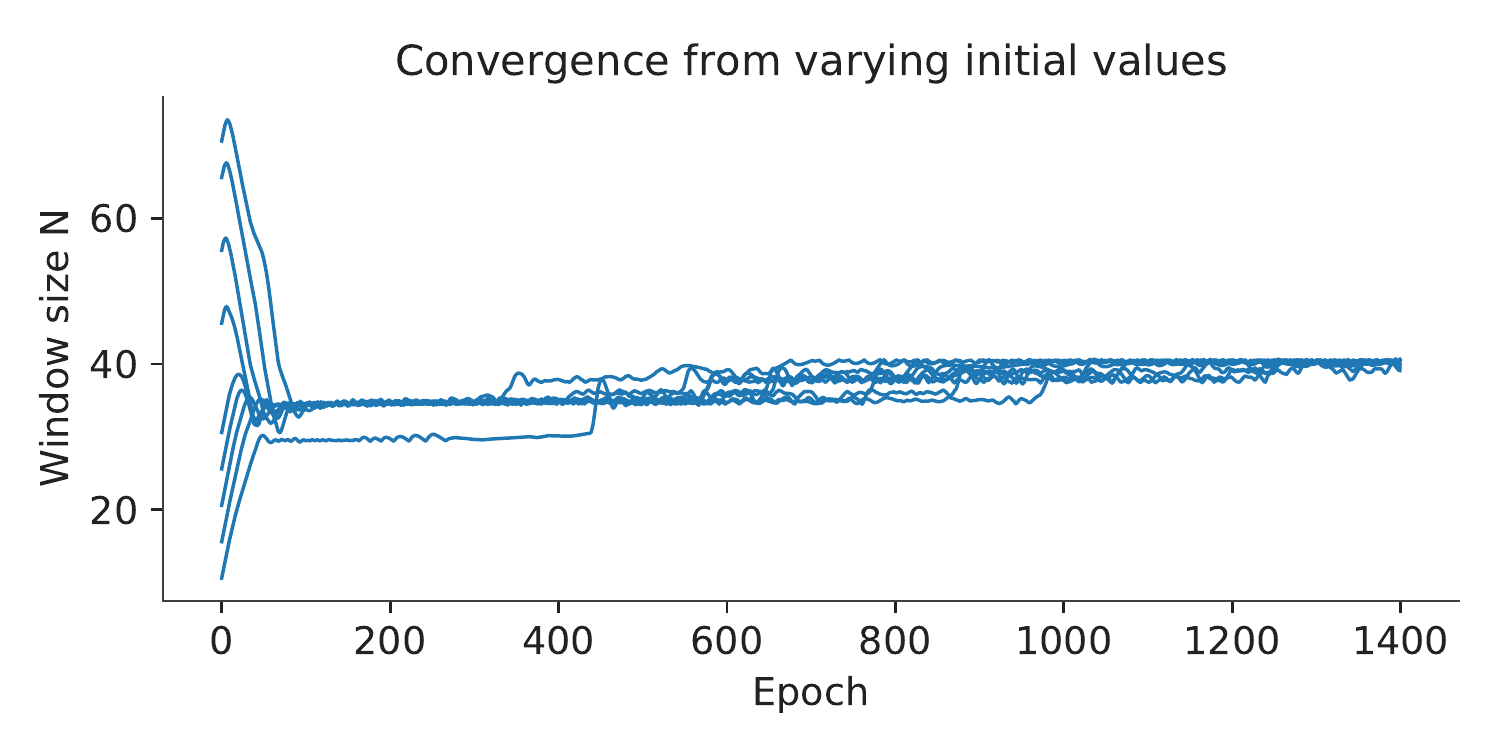}
	\caption{In this plot we show the convergence of the window size estimation as we start from different initial values.  Note that we get consistent results and fast convergence.}
	\label{fig::Nepoch}
\end{figure}

Let us consider a simple input $x[m]$ which consists of 2 alternating sinusoids of frequencies $f_1$ and $f_2$, both having the same length, $L$. We want to build a classifier which when given input STFT frames of the signal $x[m]$ would classify them as being $f_1$ or $f_2$ respectively. We know that to discriminate these signals from their spectra, we need to consider a window length of $N \leq L$, because any window length greater than $L$ would smear the spectra over time, which would make it tough for the classifier to distinguish the two classes. Likewise, for very short windows we will observe increasing smearing along the frequency axis which can also impede classification (examples of each case can be seen in \autoref{fig::class_spec}). However, if $N = L$, then the obtained spectral frames contain sufficient information to successfully discriminate the input signals. The more general case of this problem is a common issue when performing sound classification (or other tasks in the time/frequency domain), since the choice of the STFT parameters can significantly bias the results.  We use this simple example as a simple illustration of this problem with no loss of generality.
Instead of picking $N = L$ a priori (which in practice we would not know), we obtain a gradient for $N$ by differentiating $\mathcal{L}_C$, and then let this model figure out what the optimal $N$ should be as it simultaneously adjusts the classification parameters. We verify that this indeed behaves as expected by a simple experiment, by starting the optimization from multiple initial values of $N$ and observe that they quickly converge to the optimal value of $N=40$ as the classifier is being trained (Figure \ref{fig::Nepoch}).

At this point we need to make a very important observation. Had we wanted to perform an exhaustive search for the optimal value of $N$ we would have to retrain the classifier for each choice of $N$. Instead, by jointly optimizing both $N$ and the classifier we vastly reduce the amount of forward passes that we need to perform. For this experiment the classifier will converge after about 1500 iterations, regardless of whether we use a fixed $N$, or one that is concurrently optimized.  In effect, we sped up the search for an optimal $N$ by a factor as big as the original search points.  Given that today we often work with systems that can take days to train, such a speed up can be significant.

\section{Optimizing for dynamically changing parameters}
Often, an input signal changes over time and that necessitates changing the STFT parameters in response. For example, a signal might exhibit low-frequency elements that move slowly, which suggests a long analysis window, but at a later segment it might contain short-term events which necessitate shorter analysis windows. In this section we will show how one can optimize for a continuously changing window (or hop) size using gradient descent.  We will use again the sparsity cost as above, but this time the input signal will necessitate different settings at varying times.  Instead of obtaining one parameter value that is globally optimal, we will instead produce a set of locally optimal values resulting in an STFT with a dynamically changing analysis window.  In order to achieve such dynamically distributed windows, our framework needs to have three degrees of freedom: number of windows, length of each window, and overlaps between windows.

To accommodate that, we introduce the idea of a \textit{mapping function}, which maps the index of each window to its corresponding sample position in the input sound.  For example, for equally spaced windows this function would be a simple linear relation between the order index of a window and on which input time index that window is centered.  So the first window (order index $i=0$) would be centered at sample index $m=0$, whereas the $k$'th window will be centered in $m = h\cdot k$, where $h$ is the STFT hop size.  If its slope of this relationship gets steeper (larger $h$), the windows become longer and more sparsely distributed, whereas a shallower slope (smaller $h$) will result in closely packed windows.  If this mapping function is not a straight line, then the windows will not be uniformly distributed which is what we will use in this section.

\begin{figure}[ht]
\centering
\includegraphics[width=0.47\textwidth]{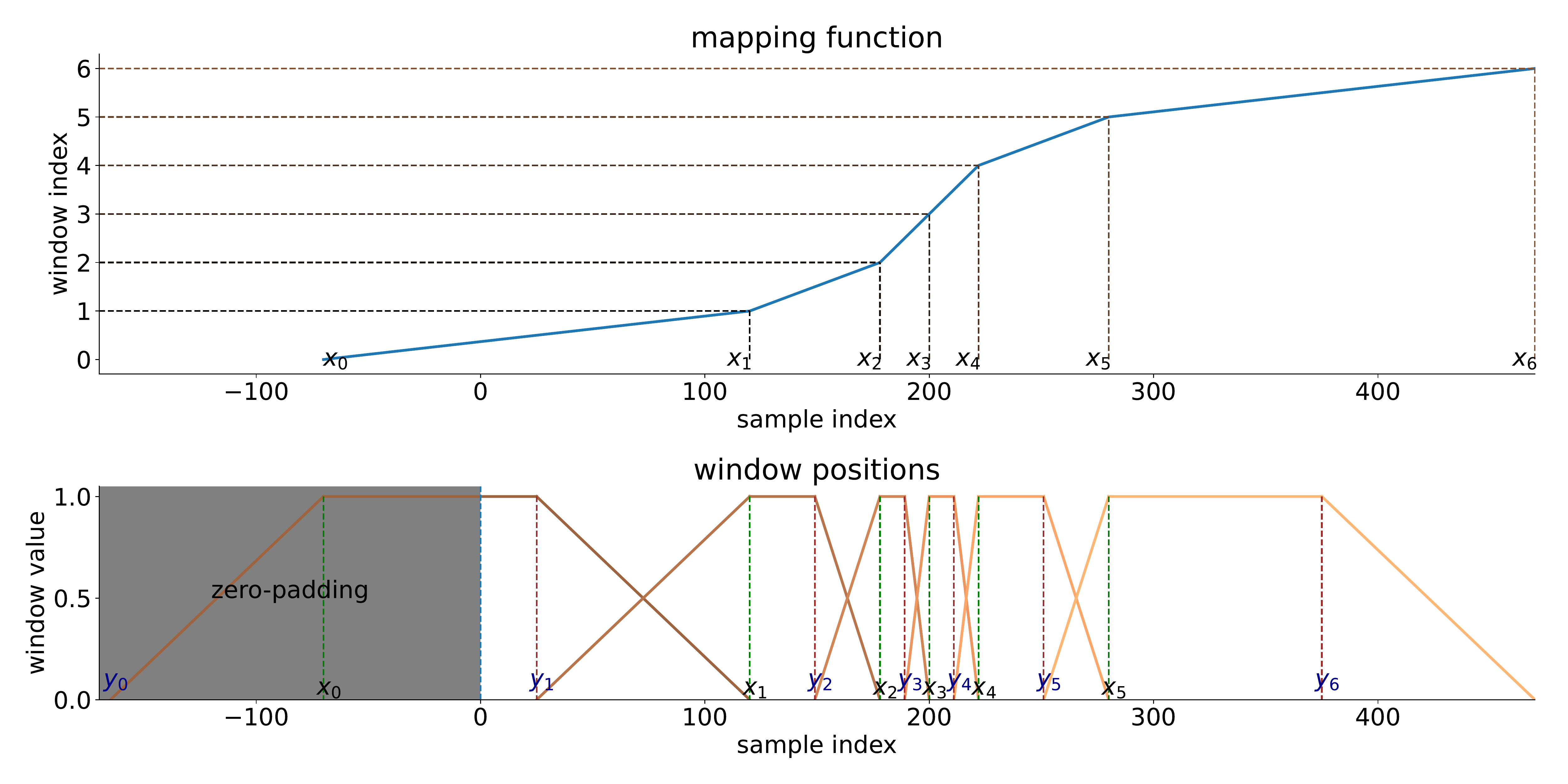}
\caption{Example mapping function and corresponding trapezoid windows.  The top plot shows the mapping function that translates a window index to a sample index.  The implied trapezoid windows from this function are shown in the bottom plot.  We see that each successive window starts at the index given by the mapping function.  In addition to that, the slopes of the trapezoids are such so that all windows sum to one, ensuring a proper sampling of the input signal.}
\label{mapf}
\end{figure}

In this setting, $W_m[n]$ from the parameterized STFT in equation \ref{stft} will contain in it such a mapping function that will map each window to an arbitrary location of our input signal.  We will use the trapezoid window as an example for our adaptive STFT since this will allow us to incorporate a variable hop size while ensuring a constant overlap between our windows \cite{SASPWEB2011}.  Depending on one's constraints, other window formulations are also possible.  Our trapezoid window function is defined as:
\begin{equation}
\begin{split}
\begin{aligned}
    \mathcal{T}(m, x_i, y_i, \\
    x_{i+1}, y_{i+1})
\end{aligned}
    =
\begin{aligned}
    \begin{cases}
    \frac{m - y_i}{x_i - y_i} & \text{if } y_i \leq m < x_i \\
    1 & \text{if } x_i \leq m < y_{i+1} \\
    1 - \frac{m - y_{i+1}}{x_{i+1} - y_{i+1}} & \text{if } y_{i+1} \leq m < x_{i+1} \\
    0 & \text{otherwise}
\end{cases}\\
\end{aligned}
\end{split}
\end{equation}
where $x_i$ is the sample position of the window with order index $i$, representing start of flat region of the trapezoid window. $y_{i+1}$ is the sample position at the end of the flat region in the trapezoid for window index $i$, which is also the start of the slope for the next window with index $i+1$.  For each window $i$, $y_i$ and $x_{i+1}$ denote the beginning and end of the window. %is the beginning of the window and $x_{i+1}$ is the end of the window.
By using this formulation we can adjust the slopes of the windows so that they overlap-add to one. The relationship between the mapping function and trapezoid windows is illustrated in Figure \ref{mapf}. Note that the mapping function only needs to encode the $x_i$, we get all the $y_i$ via a different mechanism \footnote{We do not encode the $y_i$ in the mapping function so that we can facilitate other types of window parameters which might not be location-based, e.g. $y_i$ could have been a window length factor.}.

\begin{figure*}[t]
  \centering
  \begin{subfigure}[b]{0.48\textwidth}
    \includegraphics[width=\textwidth]{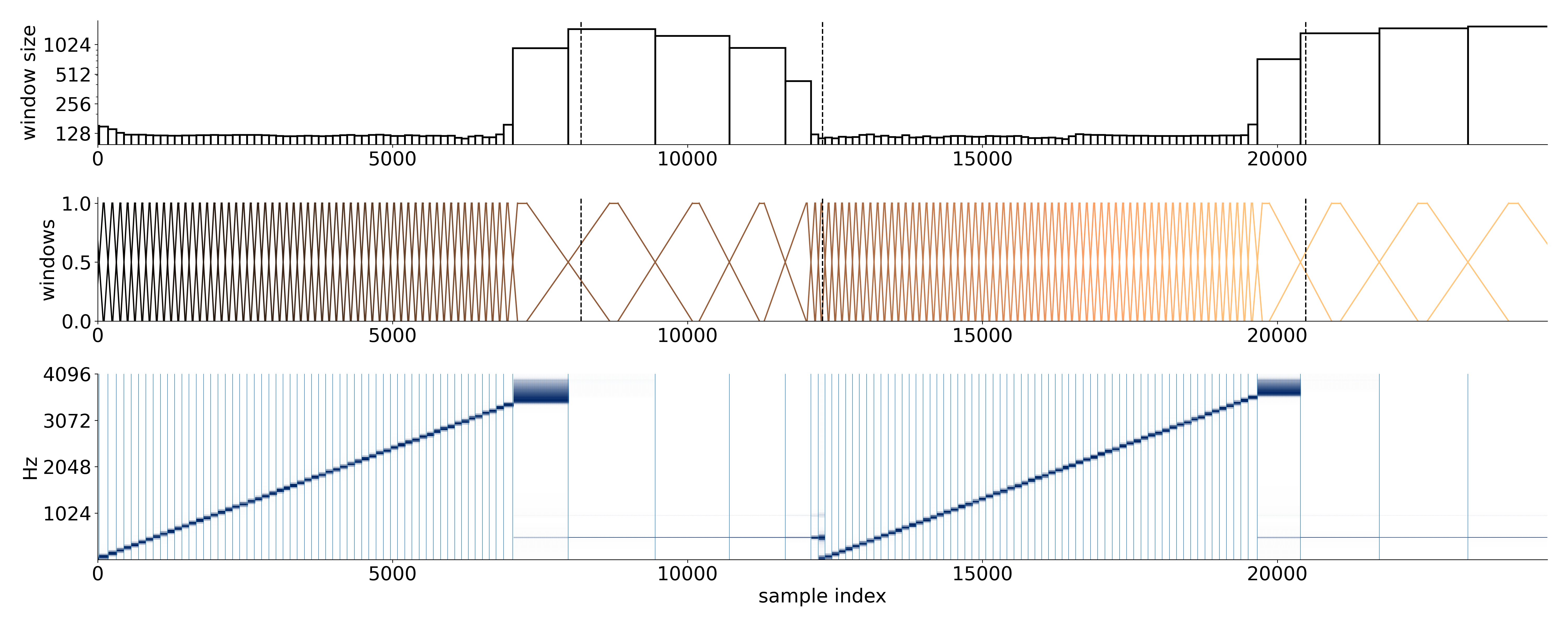}
  \end{subfigure}
  \quad
  \begin{subfigure}[b]{0.48\textwidth}
    \includegraphics[width=\textwidth]{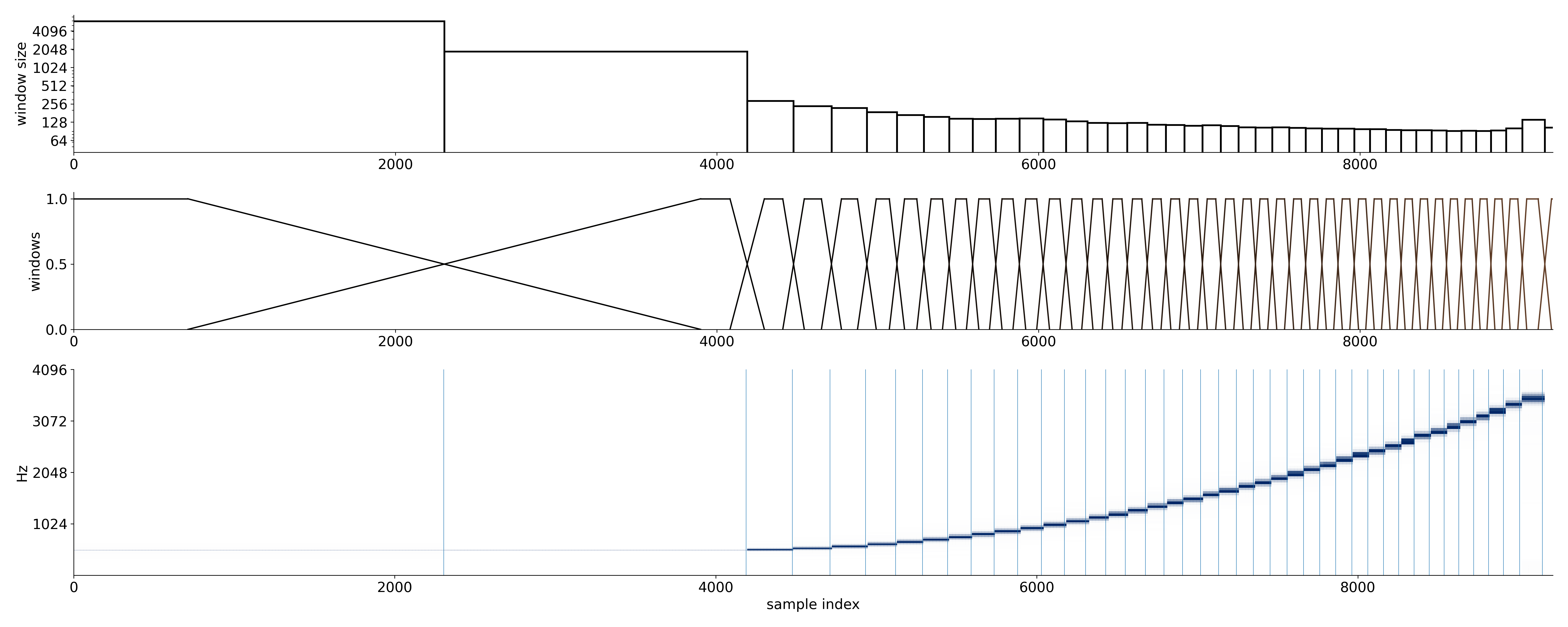}
  \end{subfigure}
  \caption{Learning an adaptive STFT on example signals.  On the left the input is a signal alternating between a chirp and a sine, and on the right an exponential chirp.  The top plots show the estimated sizes of each analysis window with respect to its position.  The middle plots show how the windows were applied, and the bottom plots show the resulting analysis with dynamically changing parameters.}
  \label{d12}
\end{figure*}
Since the estimation of a flexible mapping function should not be constrained by a simple parametric form, we use Unconstrained Monotonic Neural Networks (UMNN) \cite{UMNN} to represent it.  UMNNs produce monotonic functions by integrating neural networks with strictly positive outputs, aligned with our expectation of windows which are ordered from left to right. Using this formulation we can represent arbitrary mapping functions while maintaining the ability to differentiate the entire process. Since there is no enforced constraint that the first and last windows will be perfectly positioned at the start and end of the input signal, in practice we estimate the mapping function and zero pad the ends accordingly to facilitate windows that extend past the range of our input.

We would also like this system to be able to freely push windows out and squeeze windows in from both ends of the signal. To achieve that, we map the zero window index to the center of the input sequence (as opposed to the start). That allows us to use a mapping function that is free to push windows in and out the ends of the signal by freely manipulating the map on both sides.  Had we used a formulation that clamped the zero window index at the start of the signal we wouldn't be able to introduce new windows in the beginning and would constrain our optimization. This is more of an implementation detail, and does not change the UMNN model since it simply involves reinterpreting the window index. While the UMNN provides us with estimates the $x_i$ for each window, we use a simple feed-forward neural network that processes these $x_i$ and provides as an output their corresponding $y_i$.  This makes the entire process fully differentiable, and once we define a cost function we can now directly optimize dynamically changing window and hop sizes.
%To avoid local minima and get a reasonable window distribution we add an , a certain constraint is needed beyond cost function.  We are optimizing the sum of local concentration. However, the range of the value of concentration could be very large, so we clamped the value $C[i, W]$ to norm of all the local concentration. Note that the absolute value of loss does not mean much in this case due to clamping. 
%\begin{equation}
%\begin{aligned}
%    N &= \sqrt{\sum_{i} \mid C[i,W] \mid ^2}\\
%    \mathcal{L}_{i} &= 
%    \begin{cases}
%    C[i,W]& \text{if } N\geq C[i, W] \\
%    N + 10^{-4}(C[i,W] - N)              & \text{otherwise} \\
%\end{cases}\\
%\mathcal{L} &= \sum_{i = 0}^{} \mathcal{L}_{i}
%\end{aligned}
%\end{equation}
\subsection{Sparsity Experiments}
To verify that the proposed method works, we once again optimize for a sparse STFT output as an illustration, but this time using a signal that requires different analysis settings at each part. The loss we will use is now defined as:
\begin{equation}
\mathcal{L}_q = -\sum_{\forall m} C[m, W]
\label{lq}
\end{equation}
i.e. we are simply adding the frame-wise concentrations. One last issue to address is that this particular loss function produces many local optima since different window distributions could result in similar levels of sparsity. This usually happens when a single frame dominates the summation in equation \ref{lq}. In order to address this issue we clip large values of $C[m,W]$ to the Frobenius norm of concentration of all frames, which for this particular loss function eliminates issues with local optima.

We show results from three example inputs. All these examples include a signal that locally necessitates a different window size.  Using a fixed STFT window size throughout will not result in best results in certain sections.  First we use a simple signal, consisting of an alternating chirp and a constant frequency sinusoid as shown in Figure \ref{d12}.  The chirp portions are best described by a short analysis window that captures the temporal changes, whereas the constant frequency parts are best described by a longer window that minimizes frequency smearing. By optimizing the sizes we get the results shown in the same figure.  We see that training via gradient descent added more windows in the chirp sections and used much longer windows in the sinusoid sections.
% \begin{figure}[ht]
% \centering
% \includegraphics[width=0.47\textwidth]{sine_sweep_2.png}
% \caption{Learning an adaptive STFT on a signal alternating between a chirp and a sine.  The top plot shows the size of each analysis window of the adaptive STFT with respect to its position.  The middle plot shows how the windows were applied, and the bottom plot shows the resulting adaptive window size analysis. Note that the entire structure of this analysis was estimated using gradient descent and was not hand-tuned.}
% \label{d1}
% \end{figure}

A second example is also shown in Figure \ref{d12}.  Here we have an exponential chirp, which means that as the frequency of the input rises the speed of the frequency change also grows.  This means that we will need increasingly smaller windows to properly represent the rapid change of frequency without smearing the spectral estimates over time.  As can be seen from the plots, by using gradient descent our approach results in an optimal decomposition, where the window size shrinks over time to accommodate the input's characteristics.
% \begin{figure}[ht]
% \centering
% \includegraphics[width=0.47\textwidth]{sine_sweep_trape_4.png}
% \caption{Learning an adaptive STFT on an exponential chirp. The plots use the same format as in figure \ref{d1}. One can clearly see that the resulting window segmentation is doing what is expected.}
% \label{d2}
% \end{figure}

\begin{figure}[ht]
\centering
\includegraphics[width=0.47\textwidth]{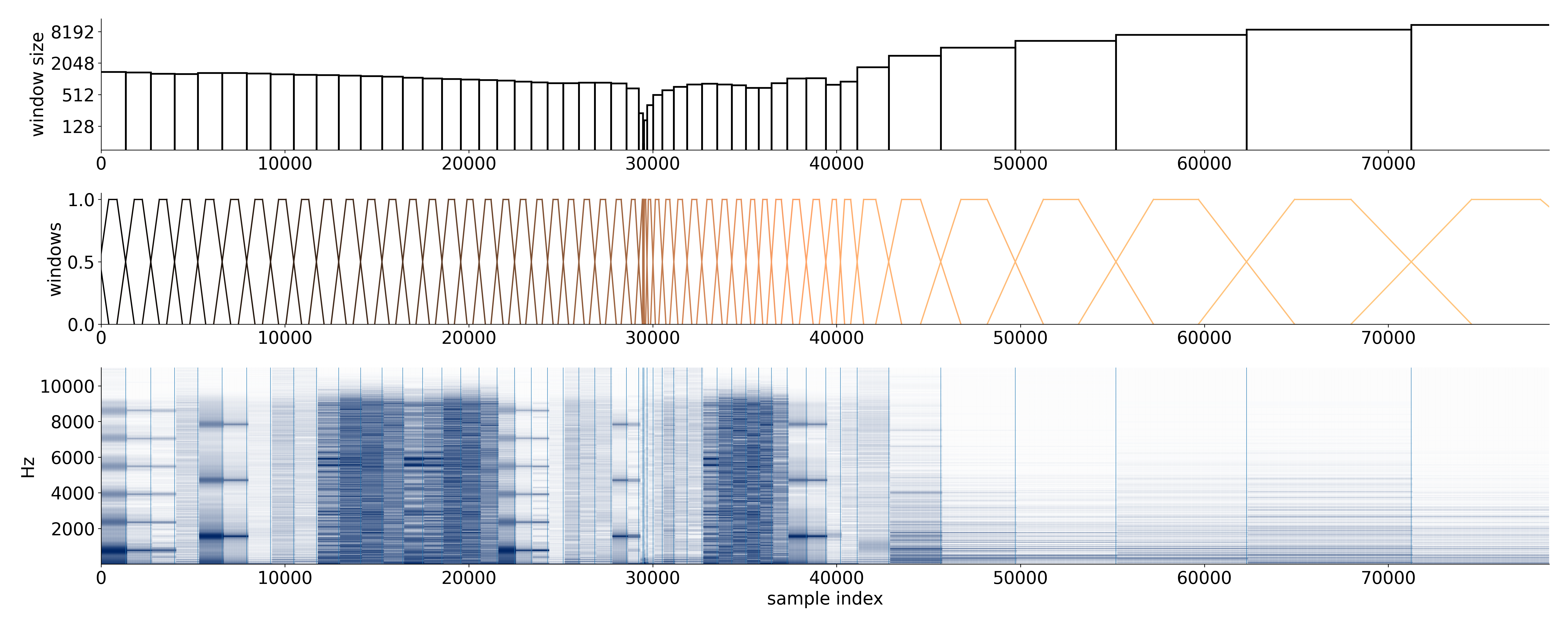}
\caption{Learning an adaptive STFT on real sounds.  The figure elements are the same as Figure \ref{d12}.  Once again we see that the resulting STFT properly adapts to the input signal to sparsely represent it.}
\label{d3}
\end{figure}

Lastly, we show a third example using real sounds. In this case we have some drum sounds in the first half and piano chords in the second half.  The section with the drum sounds requires a range of window sizes, from large to accommodate pitched drums (at the start), to short for impulsive sounds (around sample index 30,000).  The piano section requires longer windows to best describe the low sustained chords.  As shown in Figure \ref{d3} our proposed approach again finds appropriate windows for each section.
\section{Discussion}
In this paper we showed various approaches that allow us to use gradient descent optimization on the parameters of an STFT analysis. The significance of this work is that it provides a way to include STFT analysis parameters in broader optimization contexts, e.g. as trained parameters of a neural net that accepts the resulting STFT inputs. As shown in our experiments, jointly optimizing the STFT with the subsequent task at hand we can obtain the optimal STFT values with fewer evaluations than performing a search. We hope that using this approach one can incorporate the search for optimal STFT values in a global optimization setting and thus eliminate what is almost always a slower manual exhaustive search.
%\newpage
\bibliographystyle{IEEEbib}
\bibliography{strings,refs}

\end{document}